\documentclass[preprint2]{aastex701}

\usepackage{amssymb}
\usepackage{lipsum}
\usepackage{bm}
\usepackage{amsmath}
\usepackage{wasysym}
\usepackage{cancel}
\usepackage{xspace}
\usepackage{comment}

\begin{document}


\title{The Eschatian Hypothesis}

\author[orcid=0000-0002-4365-7366,sname='Kipping']{David Kipping}
\affiliation{Columbia University, 550 W 120th Street, New York NY 10027, USA}
\email[show]{dkipping@astro.columbia.edu}

\begin{abstract}
The history of astronomical discovery shows that many of the most detectable phenomena, especially detection firsts, are not typical members of their broader class, but rather rare, extreme cases with disproportionately large observational signatures. Motivated by this, we propose the Eschatian Hypothesis: that the first confirmed detection of an extraterrestrial technological civilization is most likely to be an atypical example, one that is unusually ``loud'' (i.e., producing an anomalously strong technosignature), and plausibly in a transitory, unstable, or even terminal phase. Using a toy model, we derive conditions under which such loud civilizations dominate detections, finding for example that if a society is loud for only $10^{-6}$ of its lifetime, it must emit ${\gtrsim}1$\% of its total observable energy budget during that phase to outrun quieter populations. The hypothesis naturally motivates agnostic anomaly searches in wide-field, multi-channel, continuous surveys as a practical strategy for a first detection of extraterrestrial technology.
\end{abstract}



\section*{}

Observational bias is a powerful and familiar force in astronomy. The phenomena we most easily detect are, quite often, not particularly common. As an example, the first exoplanets that were assuredly discovered were found around a pulsar of all places, PSR 1257+12 \citep{wolszczan:1992}. This challenged the earlier expectation that exoplanets would likely resemble those found in the Solar System \citep{lissauer:1993}. Indeed, the reality of these exoplanets was openly questioned in the astronomical community \citep{gil:1993}. According to the NASA Exoplanet Archive (NEA) as of 3rd December 2025, amongst the 6052 confirmed exoplanets, only eight were found around pulsars and we thus now understand them to be highly atypical examples. This lesson was repeated when the radial velocity technique started detecting exoplanets around normal stars, beginning with 51 Pegasi\,b \citep{mayor:1995}. Before 2000, 22 exoplanets were confirmed and seven of those have semi-major axes $<0.1$\,au and masses exceeding that of Saturn (see NEA; \citealt{NEA:2025}) - what we'd typically label as a hot-Jupiter. But rather than representing a third of all exoplanets, we now understand that they are incredibly rare, with less than one percent of Sun-like stars hosting such worlds \citep{wright:2012,fressin:2013}.

And of course, over-representation of unusual astronomical phenomena in our surveys is not limited to exoplanetary science. One merely needs to look up at the night sky to note that approximately a third of the naked-eye stars are evolved giants, despite the fact less than one percent of stars are in such a state - a classic observational effect known as Malmquist bias \citep{malmquist:1922}. Or consider that a supernova is expected roughly twice per century in Milky Way-sized galaxies \citep{tammann:1994} - an astoundingly rare event. And yet, despite being an inherently rare type of transient, astronomers routinely detect thousands of supernovae every year \citep{nicholl:2021}, as a product of their enormous luminosities.

If history is any guide, then perhaps the first signatures of extraterrestrial intelligence will too be highly atypical, ``loud'' examples of their broader class. Leaning into this analogy further, consider that giant stars and supernovae are transitory - representing relatively brief periods of their total lifetimes. These episodes are triggered by an abrupt change in hydrostatic equilibrium, and in both cases are highly unsustainable. For a civilization comparable to our own, the brightest luminosity we could achieve would be a global nuclear war\footnote{Detonation of the world's nuclear stockpile was in fact suggested as a METI strategy by James Elliot in a 1971 SETI meeting \citep{charbonneau:2024}.} - which too is obviously unsustainable. The ``Eschatian Hypothesis'', outlined here\footnote{Derived from the ancient Greek ``eschatos'', meaning ``last'' or ``final'', for example eschatology studies the destiny of humanity.}, thus argues that humanity's first confirmed detection of another intelligence could be that of an inherently unstable, transitory, atypical but very loud example.

The notion that loud civilizations are inherently unstable is not strictly required to manifest the outlined observational bias, but it is well-motivated via the ``Sustainability Solution'' \citep{haqq:2009} and the concept of ``Great Filters'' \citep{hanson:1998}. Indeed, examples of humanity's non-sustainable practices have already been suggested as possible technosignatures, such as anthropogenic climate change \citep{sheikh:2025}, ozone-destroying CFCs \citep{haqq:2022} and atmospheric pollution \citep{kopparapu:2021}. More generally, all technosignatures represent some kind of departure from natural equilibrium.

It remains unproven that the (presumably) rare examples of loud civilizations will truly be more likely to be detected, against the backdrop of what is (presumably) a more abundant quiescent population. To explore this, I present a very simple toy model where civilizations have an average lifetime of $T$ and can be split into two groups, with a fraction $f_+$ being loud and $f_-(=1-f_+)$ being quiet. The quiet ones remain quiescent with luminosity $L_-$ (radiated power in whatever detection channel one is interested in; see \citealt{wright:2014}) their entire lifetime, $T$. The loud ones have luminosity $L_+$ for a fraction $t_+$ of their lives and are thus quiet ($L_-$) for the remaining $t_-(=1-t_+)$ fraction. Detectability scales as $L^{3/2}$ (where $L$ is the effective luminosity in the chosen technosignature channel), since the observable volume grows as the radius cubed and radius goes as the square root of luminosity via the inverse-square law. On top of this, the actual occurrence goes as $f_{\pm} t_{\pm}$. Accordingly, the number of discoverable loud civilizations, versus quiet ones, will be

\begin{align}
\frac{N_+}{N_-} &= \frac{ f_+ t_+ L_+^{3/2} }{ f_+ t_- L_-^{3/2} + f_- L_-^{3/2} },
\end{align}

where the denominator includes the contribution of both persistently quiet civilizations and the loud ones merely in a quiet phase. The above simplifies to

\begin{align}
\frac{N_+}{N_-} &= \Big(\frac{ f_+ t_+ }{ 1 - f_+ t_+ }\Big) \Big(\frac{L_+}{L_-}\Big)^{3/2},\nonumber\\
\qquad&\simeq f_+ t_+ (L_+/L_-)^{3/2} = \eta_+ (L_+/L_-)^{3/2},
\label{eqn:Nratio}
\end{align}

where I have assumed $f_+ t_+ \ll 1$ and used $\eta_+ = f_+ t_+$ (duty-cycle). Thus, loud civilizations are likely detected first when $L_+/L_- > \eta_+^{-2/3}$. For example, if the duty cycle of civilizations being loud is $\eta_+ = 10^{-6}$, then the Eschatian Hypothesis holds if $L_+/L_- > 10^{4}$.

One can also write down a causal model between $t_+$ and $L_+$ governed by energy balance, where one should expect that as $t_+ \to 1$ then $L_+ \to L_-$ (i.e. $L_+$ decreases as $t_+$ increases). Consider that both quiet and loud civilizations have an average accessible energy of $E$ over their lifetimes. Loud civilizations release an energy $\alpha E$ during their loud phase, and thus loud civilizations have a luminosity $L_+ = (\alpha E)/(t_+ T)$. By comparison, one can write that the quiet luminosity\footnote{
To be pedantic, civilizations that are quiet for their entire history have a luminosity $E/T$, whereas loud civilizations have a quiet-phase luminosity of $(1-\alpha)E/((1-t_+) T)$. However, assuming $\alpha\ll1$ and $t_+\ll1$, these are approximately equal.
} is $L_- \simeq E/T$. Thus, the luminosity ratio follows

\begin{align}
\frac{L_+}{L_-} &\simeq \alpha/t_+.
\label{eqn:Lratio}
\end{align}

Note that by definition, $\alpha > t_+$ (else they'd be quieter during the loud-phase). Plugging Equation~(\ref{eqn:Lratio}) into Equation~(\ref{eqn:Nratio}), one obtains

\begin{align}
\frac{N_+}{N_-} &\simeq f_+ \alpha^{3/2} t_+^{-1/2}.
\end{align}

So for loud civilizations to dominate, one requires $\alpha > (t_+/f_+^2)^{1/3}$. For example, in the limit where all civilizations go loud ($f_+\to1$), then if civilizations spend $t_+ = 10^{-6}$ of their lifetime in a loud state (e.g. 4 days out of a 10,000 year history), then one requires them to release 1\% of their lifetime energy budget out during this phase.

In practical terms, the Eschatian Hypothesis suggests that wide-field, high-cadence surveys optimized for generic transients may offer our best chance of detecting such loud, short-lived civilizations. Facilities such as the Vera C.~Rubin Observatory, EvryScope, PANOPTES, and the Gaia alert stream are already moving toward a regime where the sky is effectively monitored as a time-domain dataset. Rather than targeting narrowly defined technosignatures, Eschatian search strategies would instead prioritize broad, anomalous transients -- in flux, spectrum, or apparent motion -- whose luminosities and timescales are difficult to reconcile with known astrophysical phenomena. Thus, agnostic anomaly detection efforts (e.g. \citealt{giles:2019}; \citealt{wheeler:2019}) would offer a suggested pathway forward.

\section*{Acknowledgements}

I acknowledge earlier helpful conversations at a PSETI Meeting with Ravi Kopparapu on observational bias in a SETI context.

\newpage



\end{document}